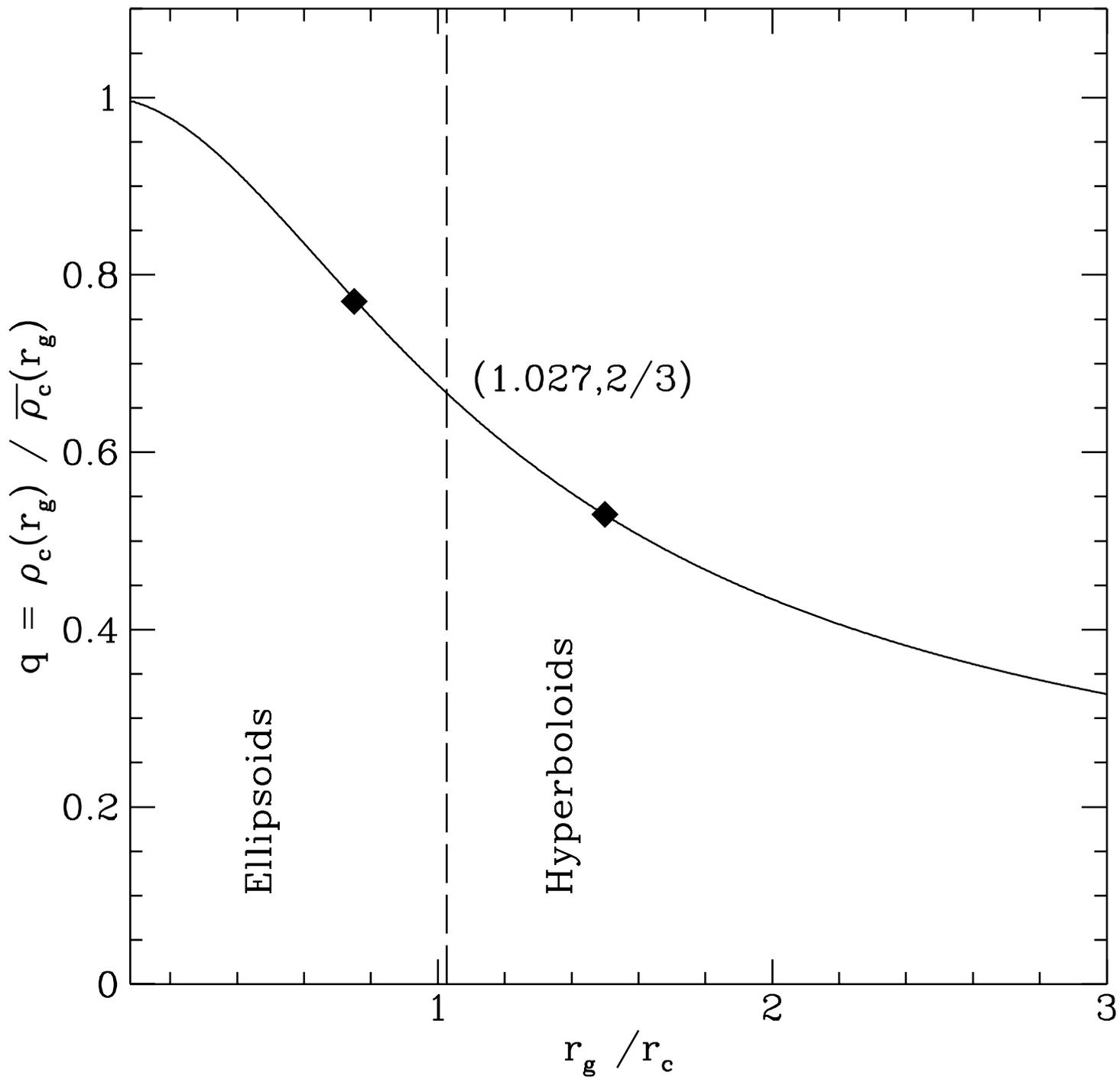

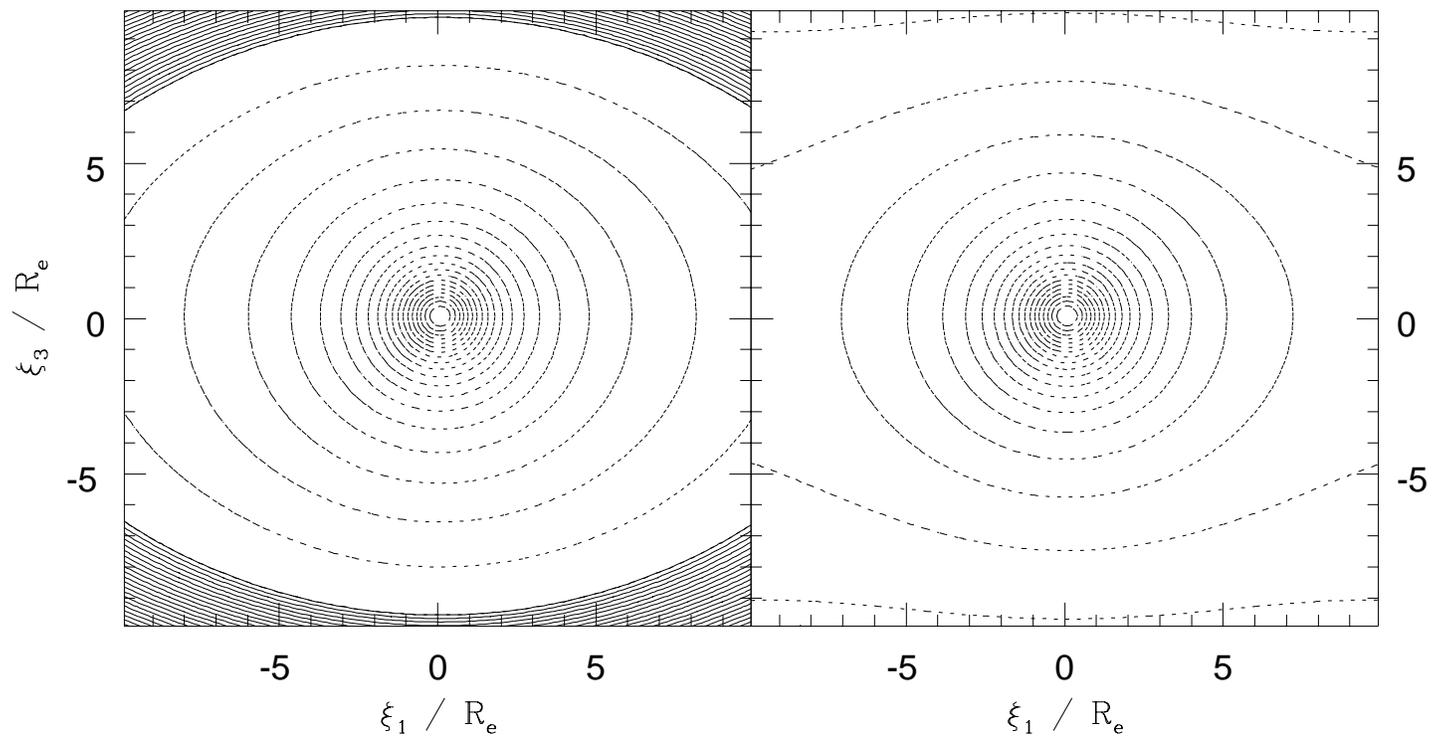

# Alignment and morphology of elliptical galaxies: the influence of the cluster tidal field

L. Ciotti[1,2] and S.N. Dutta[2]

[1] *Osservatorio Astronomico di Bologna, via Zamboni 33, Bologna 40126, Italy*
[2] *Princeton University Observatory, Peyton Hall, Princeton 08544, NJ, USA*

**ABSTRACT**
We investigate two possible effects of the tidal field induced by a spherical cluster on its elliptical galaxy members: the modification of the ellipticity of a spherical galaxy and the isophotal alignment in the cluster radial direction of a misaligned prolate galaxy. Numerical N-body simulations have been performed for radial and circular galactic orbits. The properties of the stars' zero–velocity surfaces in the perturbed galaxies are explored briefly, and the adiabaticity of the galaxy to the external field is discussed. For a choice of parameters characteristic of rich clusters we find that the induced ellipticity on a spherical galaxy is below or close to the detectability level. But we find that the tidal torque can result in significant isophotal alignment of the galaxies' major axis with the cluster radial direction if the galaxy is outside the cluster core radius. The time required for the alignment is very short compared with the Hubble time. A significant increase in the ellipticity of the outer isophotes of the prolate model is also found, but with no observable isophotal twisting. Our main prediction is an alignment segregation of the elliptical galaxy population according to whether their orbits lie mostly outside or inside the cluster core radius. These results also suggest that galactic alignment in rich clusters is not incompatible with a bottom-up galaxy formation scenario.

**Key words:** galaxies: clustering – evolution – interactions

## 1 INTRODUCTION

The morphology of elliptical galaxies (Es) is used to a considerable extent to acquire information about their structure and evolution. This is justified, to the first approximation, by the fact that their internal two body relaxation time is longer than the Hubble time, so the shape and the internal dynamics of Es probably reflect the conditions immediately after the end of their formation processes.

The problem with this naïve scenario is that Es in clusters are not isolated objects, and are subject to the effects of their environment, many of which act on time scales shorter than the Hubble time. Consequently it is of utmost importance to be able to isolate the effects of the environment when attempting to decipher the initial conditions from the present. Also these effects are of intrinsic interest as a source of information of the environment. In some cases the past interactions are signalled by the presence of shells, counter-rotating cores, polar rings, dust lanes, etc., which are taken to be evidences for merging events in the history of the galaxy. In other cases we know that strong environmental effects have to be present, for example in interacting galaxies and galaxies in groups.

Another group of interactions are the hyperbolic encounters between galaxies in rich (well relaxed) clusters, and the effect of the mean cluster tidal field (CTF), both of which are poorly understood. In particular the CTF acts continuously, and affects more or less all the galaxies in the cluster. So we expect that, even if the CTF is not as strong as other interactions, it is possible to detect its effects through systematic (albeit small) differences between cluster and field galaxies. Related to this is the cosmological problem raised by a possible large scale alignment of Es major axis along some preferred direction in the parent cluster: alignment is viewed as a consequence of top-down galaxy formation scenarios because it is usually believed that in a bottom-up cluster formation Es would be characterized by random orientation (see Djorgovski 1987 for a review).

There have been a large number of photometric studies to search global correlations of Es' properties such as ellipticities or the position angles of their major axis with their position or orientation with some particular direction in their own parent cluster. As far as any systematic differences between cluster and field Es are concerned, the results are generally in the negative. Thompson (1976) does not find any difference between the ellipticity distributions of Coma Es and field ellipticals; Fasano & Bonoli (1989) studied the occurrence of isophotal twisting in isolated galaxies, and found essentially the same fraction of isophotal twist as in cluster galaxies. Lambas, Maddox & Loweday (1992), in a homogeneous and very big sample, find that "the projected distributions of axial ratios of high and low local density environment ellipticals are remarkably similar". So they "see no evidence in the data that the environment has significantly affected the shapes of Es". Porter et al. (1991) study the Brightest Cluster Ellipticals (BCE) in elongated Abell clusters and do not find correlation between BCE shape and environment or cluster properties.

Moving to the alignment problem, there are no firmly established results. Hawley & Peebles (1975) reported "a possible indication that the galaxies are preferentially aligned along the radius vector to the center of the cluster" for the Coma cluster, a result confirmed by Thomson (1976), and suggested the CTF as a possible cause in aligning the galaxies. MacGillivray & Dodd (1979a,b) found that "there are systematic alignment tendencies for galaxies which are members of the loose cluster centered on the galaxy NGC 439 in the sense that the galaxies are preferentially aligned either parallel with the radius vector from NGC 439 or perpendicular to it". For another rich and symmetrical cluster they found a preferential alignment of Es' major diameters in the cluster radial direction. Adams, Strom & Strom (1980) find in 7 very elongated clusters, a general trend for Es to be aligned with the cluster major axis, and for the cluster A2197 they confirm the results of Thompson (1976). They also find a small but significant number of Es with their major axis perpendicular to the cluster major axis. A moderately significant result of Es radial alignment is reported by Fong, Stevenson & Shanks (1990), together with a smaller positive result for Es to be oriented perpendicularly to this direction. The best evidence for alignment is for the BCE. Trevese, Cirimele & Flin (1992) found "a strong alignment of the brightest galaxy major axis with the long axis of the parent cluster" but "no significant alignment for galaxies fainter than the first two". In this particular case the alignment is believed to be due to accretion. Rhee & Roos (1990) performed N-body simulations in which they found the BCE alignment being effected during the cluster collapse.

This paper is devoted to the study of the influence of the CTF of a rich cluster in determining the properties of its member Es. The first of the two effects studied is the change of the ellipticity of the isophotal contours of the galaxies with increasing distance from their center, and the second is the isophotal alignment of a prolate galaxy in the cluster radial direction. The cluster is treated as an externally imposed homogeneous, spherically symmetric stationary field and we evolve, using an N-body code, spherical and prolate galaxies for many galactic dynamical times in circular and radial orbits about the cluster center. In Section 2 we describe in detail the properties of the cluster and the spatial and projected properties of the galaxy models for the initial conditions we use for the simulations. In Section 3 we derive analytically the expressions for the CTF and we study the adiabaticity of the response of spherical galaxies in radial orbit and of prolate galaxies misaligned with the cluster radial direction to the tidal field. Galaxies that are in orbits that sample the most interesting parts of the

**Figure 1.** The solid line is the radial trend of the function $q = q(r_{\rm g}/r_{\rm c})$. The vertical dashed line separate the two regions of the cluster that correspond to qualitatively different shapes of the tidal potential. The two black diamonds are the positions of the models calculated in circular orbits and the apocenters of radial orbits.

CTF according to this analytical study are evolved numerically. In Section 4 the determination of the numerical initial conditions is briefly explained, and in Section 5 we show the results of the numerical simulations. In Section 6 we present our conclusions.

## 2  THE MODELS

### 2.1  Cluster Characteristics

For the smoothed, zeroth-order density distribution that describes the cluster used in our simulations we assume the distribution given by King (1972) to fit the Coma cluster:

$$\rho_{\rm c}(r) = \frac{\rho_{\rm co}}{(1 + r^2/r_{\rm c}^2)^{3/2}}, \tag{1}$$

where $r = \|{\bf r}\|$ is the distance from the cluster center, $r_{\rm c} = 171\,{\rm kpc}$ is the "core" radius, and the distribution is truncated at some $r_{\rm t}$. We derive the central density from the well-known formula $\rho_{\rm co} = 9\sigma_{\rm c}^2/4\pi G r_{\rm c}^2 \simeq 6.4\,10^{-3}\,M_\odot/{\rm pc}^3$ (e.g. Binney & Tremaine 1987, hereafter BT87, p. 228), assuming the cluster central velocity dispersion $\sigma_{\rm c} = 1.061\,10^3\,{\rm km\,s}^{-1}$. The assumption of a spherical cluster is a very idealized one, but is justified by the fact that we want to model galaxies on which a stationary cluster field has acted for a long time, so we are considering relaxed clusters, which are characterized by spherical symmetry. The same cluster density distribution (although with different parameters) is used for example by Valluri (1993) to investigate (without taking into account the self-gravity of the galaxy) the response of a disk galaxy to the CTF.

### 2.2  Spherical Galaxy Characteristics

In order to study the effect of the CTF in deforming the shape of Es, we assume as initial conditions for the N-body simulations a Plummer galaxy model, described by the density-potential pair:

$$\rho(\xi) = \frac{3}{4\pi}\frac{M_{\rm g}}{r_{\rm P}^3}\frac{1}{(1+\tilde\xi^2)^{5/2}}, \tag{2}$$

**Figure 2.** Sections of the stars' zero-velocity surfaces in the $\tilde{\xi}_1 - \tilde{\xi}_3$ plane, in the model galaxy (2) in radial orbit. In the left panel the sections for $\widetilde{r}_g = 0.75$ ($q \simeq 0.77$) are shown, as are those for $\widetilde{r}_g = 1.5$ ($q \simeq 0.53$) in the right panel. Dashed and solid lines refer to negative and positive potential (galaxy + cluster) values respectively.

$$\phi_P(\xi) = -\frac{GM_g}{r_P}\frac{1}{\sqrt{1+\tilde{\xi}^2}}. \qquad (3)$$

where $M_g$ is the total mass of the galaxy, $r_P$ is the core radius of the Plummer sphere, and $\tilde{\xi} = \|\xi\|/r_P$ is the distance from the center of mass of the galaxy expressed in units of $r_P$. To quantify the galactic scale length on which the deformation can be observed we use the projected density profile of the unperturbed galaxy:

$$\Sigma(R) = \frac{M_g}{\pi r_P^2(1+\widetilde{R}^2)^2}, \qquad (4)$$

where $\widetilde{R} = R/r_P$ is the normalized projected radius, and from that, assuming a constant mass-to-light ratio, we derive

$$\mu(R) - \mu(0) = 5\log(1+\widetilde{R}^2), \qquad (5)$$

where $\mu(R) \equiv -2.5\log\Sigma(R) + const$. For example $R = 10 r_P$ corresponds to a 10 mag. fading of the projected surface brightness. Integrating (4) on a circular aperture one obtains the expression for the cumulative projected mass inside $R$, and the effective radius $R_e = r_P$. In order to determine a galaxy model we have to assign realistic values for the parameters in (2). We choose $M_g = 10^{12}\, M_\odot$ and $r_P = 5.0$ kpc, so the model represents a giant elliptical. The order of magnitude estimate for the half mass radius dynamical time is $t_{dyn} \equiv 1/\sqrt{G\bar{\rho}} \simeq 2.3\, 10^7$ yr, where $\bar{\rho} = 1.5 M_g/4\pi r_h^3$ is the mean density of the galaxy inside the half mass radius, $r_h = r_P/\sqrt{2^{2/3}-1}$.

### 2.3 Prolate Galaxy Characteristics

For the analysis of the alignment of a non-spherical galaxy due to the torque induced by the CTF we start by assuming as initial conditions for the N–body simulations a density distribution reproducing a prolate galaxy. Its density distribution is stratified in homeoids labeled by $m^2 = x^2/r_P^2 + y^2/r_P^2 + z^2/(r_P^2\lambda)$ and $\lambda \geq 1$:

$$\rho(m) = \frac{3M_g}{16\pi\sqrt{\lambda}r_P^3} \times \begin{cases} 1, & \text{if } m \leq 1, \\ m^{-4}, & \text{otherwise.} \end{cases} \qquad (6)$$

We use (6) instead of the more natural generalization of (2) because it is easier to obtain the potential needed to compute the internal velocity dispersion (see Section 4). We choose the total mass $M_g$ and the "core" radius $r_P$ equal to the corresponding values used for the spherical model, and $\lambda = 2$. The half mass radius dynamical time is $t_{dyn} \simeq 3.3\, 10^7$ yr, the half mass homeoid being labeled by $m = 3/2$.

As in the previous case, to obtain information on the galaxy modifications we project (6) along the $y$ axis. The resulting surface density distribution is:

$$\Sigma(l) = \frac{3M_g}{16\pi\sqrt{\lambda}r_P^2} \begin{cases} [l\sqrt{1-l^2}(2l^2-1) + \text{Arcsin}\,l]/l^3, \\ \pi/2l^3, \end{cases} \qquad (7)$$

where $l^2 = x^2/r_P^2 + z^2/(r_P^2\lambda)$ and the first expression holds for $l \leq 1$. Integrating over ellipses of constant $l$ one can easily determine the effective isophote $l_e$, i.e. the isophote that contains half of the total mass. After a little algebra we have that $l_e = 3\pi/8$, i.e. the short semi-axis of the effective isophote is $\simeq r_P$. This is consistent with the longer half mass dynamical time with respect to the spherical model. The radial behavior of the surface brightness profile, for $l \geq 1$ is given by:

$$\mu(l) - \mu(0) = 2.5 \log\left(\frac{16 l^3}{3\pi}\right) \simeq 0.58 + 7.5 \log l. \tag{8}$$

**Figure 3.** Sections of the stars' zero-velocity surfaces for the model galaxy (2) in circular orbits. From left to the right the sections are respectively in the planes $\tilde{\xi}_2 = 0$, $\tilde{\xi}_3 = 0$, and $\tilde{\xi}_1 = 0$. In the three upper panels the galaxy orbit is inside the cluster core radius at $\widetilde{r}_g = 0.75$, and in the three lower ones the galaxy orbit is at $\widetilde{r}_g = 1.5$. Dashed and solid lines refer to negative and positive potential (galaxy + cluster) values respectively.

## 3   ANALYTICAL ESTIMATE OF A GALAXY RESPONSE

In this Section we derive and study the expression for the CTF and its fundamental properties as a function of the distance from the cluster center. Following that we briefly describe the effects of the CTF on a spherical galaxy in radial and circular orbit, using the zero-velocity surfaces of its stars. Finally we discuss the response of a galaxy in radial orbit and of a prolate galaxy misaligned with the cluster radial direction to the variation of the CTF, using a quantitative estimate of the galaxy adiabaticity.

### 3.1   Cluster tidal field

Let $C = (O; \mathbf{e}_1, \mathbf{e}_2, \mathbf{e}_3)$ be an inertial reference system centered on the cluster, with $\mathbf{r}$ as a generic position vector, and $C' = (O'; \mathbf{f}_1, \mathbf{f}_2, \mathbf{f}_3)$, with $\xi$ as the generic position vector, be a second reference system centered on $\mathbf{r}_g = \int \rho(\mathbf{r}; t) \mathbf{r} d^3 \mathbf{r} / M_g$, the center of mass of the galaxy. The system $C'$ is allowed to rotate with an angular velocity $\mathbf{\Omega}$ (viewed from the frame $C'$). This is the natural system to study the response of the galaxy to the CTF. Using the usual rules of transformation (see e.g. Arnold 1978), the positions, velocities and accelerations in the two frames are related as follows,

$$\begin{aligned}
\mathbf{r} &= \mathbf{r}_g + \mathcal{O}\xi, \\
\dot{\mathbf{r}} &= \mathbf{v}_g + \mathcal{O}(\mathbf{\Omega} \wedge \xi + \dot{\xi}), \\
\ddot{\xi} &= \mathcal{O}^T(\ddot{\mathbf{r}}_{\text{self}} + \ddot{\mathbf{r}}_{\text{ext}} - \mathbf{a}_g) \\
&\quad - 2\mathbf{\Omega} \wedge \dot{\xi} - \dot{\mathbf{\Omega}} \wedge \xi - \mathbf{\Omega} \wedge (\mathbf{\Omega} \wedge \xi),
\end{aligned} \tag{9}$$

where the last three terms on the right hand side are the Coriolis' Force, Euler's Force and Centrifugal Force, respectively, and $\mathcal{O}$ is the orthogonal transformation matrix (generally time-dependent) between $C$ and $C'$.

The accelerations terms in parenthesis are the fields acting on each star due to the parent galaxy, that due to the cluster and the acceleration of the galaxy center of mass, respectively. The CTF is given by the linearization of $\ddot{\mathbf{r}}_{\text{ext}} - \mathbf{a}_g$, where, using Jeans' equations, $M_g \mathbf{a}_g = \int \rho(\mathbf{r}; t)(\ddot{\mathbf{r}}_{\text{self}} + \ddot{\mathbf{r}}_{\text{ext}}) d^3 \mathbf{r} = \int \rho(\mathbf{r}; t) \ddot{\mathbf{r}}_{\text{ext}} d^3 \mathbf{r}$. Linearizing the cluster field around $\mathbf{r}_g$ and defining $r_g = \|\mathbf{r}_g\|$, one obtains

$$\frac{M_c(r)}{r^3} r_i = \frac{M_c(r_g)}{r_g^3} r_{gi} + \\ \left[\delta_{ij} \frac{M_c(r_g)}{r_g^3} + \frac{r_{gi} r_{gj}}{r_g} \frac{\partial}{\partial r_g} \frac{M_c(r_g)}{r_g^3}\right](r_j - r_{gj}), \tag{10}$$

**Figure 4.** The solid curve ($Ad_1$) refers to the radial adiabaticity of a galaxy in free fall from infinity and the dotted curve ($Ad_2$) is the tangential adiabaticity, i.e. the adiabaticity in the direction perpendicular to the cluster radial direction. The dashed curve ($Ad_{\rm al}$) is the adiabaticity of a prolate galaxy in circular orbit around the cluster center.

where the expansion is truncated to the order $O(\|\mathbf{r} - \mathbf{r}_{\rm g}\|^2)$ and $\delta_{\rm ij}$ is the Kronecker symbol. Fixing the orientation of $C'$ in such a way that $\mathbf{r}_{\rm g}$ is parallel to $\mathbf{f}_1$, the linearized external field in (9) can be written as:

$$T_{\rm i} = \mathcal{T}_{\rm ij}\xi_{\rm j} + \mathcal{R}_{\rm i}, \tag{11}$$

where $\mathcal{R}_{\rm i}$ is the $\rm i^{th}$–component of the non–inertial forces and the expression of the (symmetric) CTF tensor is:

$$\mathcal{T}(r_{\rm g}) = -\frac{4\pi G \bar{\rho}_{\rm c}}{3} \begin{pmatrix} 3q - 2 & 0 & 0 \\ 0 & 1 & 0 \\ 0 & 0 & 1 \end{pmatrix}, \tag{12}$$

where $\bar{\rho}_{\rm c} = 3M_{\rm c}(r_{\rm g})/(4\pi r_{\rm g}^3)$ and $0 \leq q(r_{\rm g}) = \rho_{\rm c}(r_{\rm g})/\bar{\rho}_{\rm c} \leq 1$ for any non increasing density distribution. One can check equation (12) by showing the trace of the tidal tensor equals $-4\pi G \rho_{\rm c}$ as required by the Poisson equation. From (1) one obtains

$$q(r_{\rm g}) = \frac{\widetilde{r}_{\rm g}^3}{3(1 + \widetilde{r}_{\rm g}^2)^{3/2} \left[ {\rm arcsh}(\widetilde{r}_{\rm g}) - \widetilde{r}_{\rm g}/\sqrt{1 + \widetilde{r}_{\rm g}^2} \right]}, \tag{13}$$

where $\widetilde{r}_{\rm g} = r_{\rm g}/r_{\rm c}$. The solid line in Fig. 1 shows the variation of $q$ with the distance from the cluster center.

We can investigate the geometry of the CTF looking at its potential, i.e. $\mathcal{T}_{\rm ij}\xi_{\rm j} = -\partial \tau / \partial \xi_{\rm i}$, with

$$\tilde{\tau} = (3q - 2)\tilde{\xi}_1^2 + \tilde{\xi}_2^2 + \tilde{\xi}_3^2, \tag{14}$$

where $\tilde{\tau}$ is the potential, $\tau$, normalized to $2\pi G \bar{\rho}_{\rm c} r_{\rm P}^2/3$. The general property of $\tilde{\tau} = const.$ surfaces is their axisymmetry about the $\tilde{\xi}_1$ axis (i.e. the direction to the center of the cluster) and symmetry with respect to reflection about the $\tilde{\xi}_2 - \tilde{\xi}_3$ plane. We distinguish two different regions in the cluster.

For $2/3 < q \leq 1$, i.e. for $1.027 > \widetilde{r}_{\rm g} \geq 0$, (Fig. 1, region to the left of the vertical dashed line) the isopotential surfaces are prolate ellipsoids with their major axis in the cluster radial direction. The tidal potential increases with $r_{\rm g}$ with the equipotential surfaces forming a family of similar ellipsoids, i.e. inside the cluster core the CTF is everywhere compressive. For $q = 1$ (in the cluster center or in the case of an homogeneous spherical density distribution) the CTF is radially compressive about the galaxy center.

For $q = 2/3$ (Fig. 1, vertical dashed line) the isopotential surfaces are cylinders with their axis along $\tilde{\xi}_1$, and in this case the CTF is radially compressive around this axis.

For $0 \leq q < 2/3$ (Fig. 1, region to the right of the vertical dashed line) the coefficient of $\tilde{\xi}_1$ is negative. The quadratic form (14) is no longer positive definite and the value of the potential can be positive or negative, without lower or upper limits. The isopotential surfaces are two-sheet and one-sheet similar hyperboloids of revolution around the cluster radial direction, for negative and positive $\tilde{\tau}$'s values, respectively. The two families are separated by the $\tilde{\tau} = 0$ asymptotic cone, the angle of which gets smaller as $q$ increases (i.e. $r_{\rm g}$ decreases). In this last radial region the CTF is expansive along $\tilde{\xi}_1$ and radially compressive perpendicular to this axis.

So the cluster core radius ($\widetilde{r}_{\rm g} = 1$) marks the transition region between two qualitatively different CTFs, and we expect this distance to be a point of bifurcation of the Es' properties. We move now to obtain some hint for the response of a spherical galaxy in radial and circular orbit in the cluster. We do this in an approximate way by adding the appropriate tidal potential to the unperturbed galaxy potential. This is clearly not a self-consistent way of analyzing the problem but in this case is not too wrong: the CTF *increases* linearly with the distance from the galaxy center while its self-gravity in the external regions *decreases* roughly as the square of this distance. So the effects of the CTF are more important in the external parts of the galaxy. Moreover we assume that the local galaxy density is produced approximately by stars near their zero-velocity surface. The stars spend the most of their orbital time near these surfaces, justifying their use.

### 3.2 Radial Galaxy Orbit

If we consider a galaxy in a perfectly radial orbit, all the non inertial terms due to the rotation vanish, i.e. $\mathcal{R}_{\rm i} = 0$ (i= 1, 2, 3) in (11), and the system $C'$ is determined by requiring that the vector $\mathbf{f}_1$ lie on the direction of $\mathbf{r}_{\rm g}$. For our models the axis ratios of the zero-velocity surfaces near the galaxy's center are very near to one, i.e. the effect of the CTF in the central galaxy regions is totally negligible. In Fig. 2 the behavior of sections of the isopotential surfaces is shown for two different positions of the galaxy in the cluster. It is evident that in both the positions the surfaces remain approximately spherical inside $5R_{\rm e}$, and only outside they become prolate in the cluster radial direction.

### 3.3 Circular Galaxy Orbit

In this case the galaxy is in circular orbit of radius $r_{\rm g}$ around the cluster center. This assumption implies that in the expression for $\mathcal{R}_{\rm i}$ the Euler's force vanish and the Jacobi integral is conserved by each star (e.g. BT87, p. 135). The coordinate system $C'$ is determined by requiring that the vector $\mathbf{f}_1$ lie on the direction of $\mathbf{r}_{\rm g}$, and that $\mathbf{f}_3$ to be parallel to $\mathbf{\Omega}$. The zero-velocity surfaces in the galaxy are given by $\psi = \tau_{\rm eff}(\xi) + \phi_{\rm P}(\xi)$, where

$$\tau_{\rm eff} = \tau(\xi) - \frac{\|\mathbf{\Omega} \wedge \xi\|^2}{2} = \tau(\xi) - \frac{\Omega^2(\xi_1^2 + \xi_2^2)}{2} \tag{15}$$

**Figure 5.** In the lower panels the isodensity contours of the projected galaxy density in the $\xi_1 - \xi_3$ plane are shown. The galaxy is in radial orbit always inside the cluster core radius, and passes through the centre 2 times during the simulation. The time is expressed in simulations units, and the space between two ticks corresponds to 4.25 kpc. The distance, $r_g$ from the center of the cluster is expressed in kpc. In the upper panels the projection of the density in the $\xi_1 - \xi_2$ plane is shown.

and $\tilde{\Omega}^2(r_g) = 2$, where the normalization factor is that in (14).

The surfaces of constant effective CTF are cylindrical hyperboloids with the generatrix parallel to $\tilde{\xi}_2$ (and so no force acts in this direction). The CTF is expansive along the $\tilde{\xi}_1$ direction, compressive in the $\tilde{\xi}_3$ direction and neutral along $\tilde{\xi}_2$ direction, no matter what distance the galaxy is from the cluster center. If $q = 1$ (constant cluster density) than the surfaces becomes planes parallel to the orbital $\tilde{\xi}_1 - \tilde{\xi}_2$ plane, and the CTF is compressive along $\tilde{\xi}_3$ and zero in the other two directions. The zero-velocity surfaces of the sum of the CTF and (3) are not longer characterized by axial symmetry around $\tilde{\xi}_1$, as in the case of a radial galaxy orbit.

In Fig. 3 the sections of the zero-velocity surfaces are shown with respect to the principal planes for the two radii marked with dark diamonds in Fig. 1. The upper panels refer to the orbit inside the cluster core radius, and the lower one to the orbit outside it. In both cases the surfaces show a more pronounced tendency towards prolatness in the cluster radial direction than in the case of radial orbit. The transition to an observable deformation is around $4R_e$.

### 3.4 Adiabaticity

If the time-scale of change of the CTF due to the galaxy motion is long compared to the galaxy dynamical time, then the assumption that the galaxy is in local equilibrium with the CTF is correct. If the reverse is the case, or the two time-scales are comparable, the assumption of equilibrium is a poor one, the shape of the galaxy at any time may be significantly different from that of the zero-velocity surfaces, and N-body simulations are required. In order to obtain quantitative information about this time-scale, we introduce a *radial* and a *tangential* characteristic tidal time:

$$\frac{1}{t_{\mathrm{Ti}}} \equiv \left|\frac{\mathrm{d}\ln|\mathcal{T}_{\mathrm{ii}}[r_g(t)]|}{\mathrm{d}t}\right| = v_{\mathrm{ff}}(r_g)\left|\frac{\mathrm{d}\ln|\mathcal{T}_{\mathrm{ii}}(r_g)|}{\mathrm{d}r_g}\right|, \qquad (16)$$

where $t_{T1}$ and $t_{T2} = t_{T3}$ are the radial and the two equal tangential tidal times, respectively, and $r_g(t)$ is the galactic radial orbit of free-fall. The expressions for a generic spherical system are:

$$\frac{1}{t_{T1}} = v_{\mathrm{ff}}(r)\left|\frac{2/r + (\alpha - 2/r)q}{q - 2/3}\right|, \qquad (17)$$

where $\alpha(r) = \mathrm{d}\ln\rho_c/\mathrm{d}r$, and

$$\frac{1}{t_{T2}} = \frac{1}{t_{T3}} = v_{\mathrm{ff}}(r)\frac{3(1-q)}{r}. \qquad (18)$$

We use

$$v_{\mathrm{ff}}(r_g) = \sqrt{2|\phi_c(r_g)|} = \sqrt{8\pi G\rho_{c0}r_c^2\,\mathrm{arcsh}(\widetilde{r}_g)/\widetilde{r}_g},$$

i.e. for simplicity a fall from infinity. With this choice we are overestimating the rapidity at which the CTF is changing in the galaxy reference system. Defining *adiabaticity* of the galaxy along $\tilde{\xi}_i$ the dimensionless number $Ad_i \equiv t_{\mathrm{Ti}}/t_{\mathrm{dyn}}$, the condition $Ad_{\rm i} \gg 1$ means that the galaxy can "adjust" its shape along $\widetilde{\xi}_{\rm i}$ to the local CTF. Moreover, one can have the galaxy adiabatic in one direction and non-adiabatic in another. In Fig. 4 the solid line refers to the radial adiabaticity and the dotted line to the tangential one. The radial adiabaticity line separates the cluster in three regions. A very central adiabatic region ($\widetilde{r}_{\rm g} \leq 0.4$), a non-adiabatic intermediate zone ($0.3 \leq \widetilde{r}_{\rm g} \leq 1.3$), and finally the external adiabatic zone. From the divergence of $Ad_1$ near the cluster core, due to the topological change of the isopotential surfaces at that point, every galaxy in radial orbit with an apocenter greater than $r_{\rm c}$ experience a non–adiabatic radial transition. On the other hand galaxies with the same parameters as our model are in local equilibrium with the tangential CTF at any distance from the center. The divergence of the adiabaticity for $\widetilde{r}_{\rm g} \simeq 1.8$ is due to the fact that at this point the radial component of the CTF presents a maximum, i.e. its derivative is zero. ¿From the definition of adiabaticity the divergence follows. From a physical point of view, a galaxy crossing this point experiences no variation of the radial tidal field. Therefore the excursion of $Ad_1$ near $r_{\rm g} = 1.8 r_{\rm c}$ is of mathematical interest only and has no important consequences for the physics of the situation.

**Figure 6.** In the lower panels the isodensity contours of the projected galaxy density in the $\xi_1 - \xi_3$ plane are shown. The galaxy is in circular orbit at $\widetilde{r}_{\rm g} = 0.75$. The time is expressed in simulations units, and the space between two ticks corresponds to 4.25 kpc. In the upper panels the projection of the density in the $\xi_1 - \xi_2$ plane is shown.

As far as the alignment problem is concerned, a rough estimate of the characteristic time of response of the prolate model in circular orbit to the torque induced by the CTF can be easily obtained. Assuming the galaxy to be a rigid body with the major axis in the plane perpendicular to the velocity of its center of mass, and not considering the non-inertial coupling effect with the other degrees of freedom, one obtains:

$$I_{22}\ddot{\theta} = \sin\theta \cos\theta \, (\mathcal{T}_{33} - \mathcal{T}_{11})(I_{33} - I_{11}), \qquad (19)$$

where the $I_{\rm ii}$ are the galaxy principal axis of inertia, and $\theta$ is the angle between the major axis and the radial cluster direction.

The two points of equilibrium, corresponding to the major axis aligned ($\theta = 0$) and perpendicular ($\theta = \pi/2$) to the radial cluster direction are respectively stable and unstable, as in our model $I_{33} - I_{11} > 0$ and from (12), $\mathcal{T}_{33} - \mathcal{T}_{11} < 0$. Equation (19) is easily solved using elliptic functions, but for our needs it is sufficient to obtain the period by its linearization:

$$\frac{1}{t_{\rm al}} = \frac{\sqrt{4\pi G \bar{\rho}_{\rm c}}}{2\pi} \sqrt{\frac{(1-q)|1-\lambda|}{1+\lambda}}, \qquad (20)$$

where $q$ is given in (13). The behavior of the adiabaticity associated to this oscillations, $Ad_{\rm al} = t_{\rm al}/t_{\rm dyn}$, is shown for our assumed galaxy and cluster parameters in Fig. 4 (dashed line). One can see that the galaxy is completely adiabatic to the alignment, i.e. a "mean" star inside it completes many orbits in one oscillations period. Of course the approximation of the galaxy by a rigid body to study the alignment is not correct. What the analysis tells us is that the alignment should be reached in very few, if not only one, oscillation periods.

**Figure 7.** In the lower panels are shown the isodensity contours of the projected galaxy density in the $\xi_1 - \xi_3$ plane. The galaxy is in circular orbit at $\widetilde{r}_g = 1.5$. The time is expressed in simulations units, and the space between two ticks corresponds to 4.25 kpc. In the upper panels the projection of the density in the $\xi_1 - \xi_2$ plane is shown.

## 4 NUMERICAL METHOD

We have shown from the previous qualitative analysis that the problem of a galaxy in the cluster tidal field is intrinsically time dependent over the major volume of the cluster, and so it is necessary to use direct N-body simulations to obtain further insight in the subject.

For the problem of galactic deformation the initial conditions were fixed by generating a N-body system from the distribution function,

$$f = \frac{24\sqrt{2}}{7\pi^3} \frac{r_P^2 (-E)^{7/2}}{G^5 M_g^4}, \tag{21}$$

where $E$, is the total energy for unit mass of an individual star and $f$ is zero for positive energy values. This corresponds to the density profile (2) with global isotropic velocity dispersion (e.g. Aarseth et al. 1974). In order to reduce the possible initial noise due to the realization of the density distribution with a finite number of particles, the model is evolved for a few dynamical times maintaining fixed its initial position in the cluster.

Generating the initial conditions for the simulations to check the alignment effect is somewhat trickier. In this case no analytical inversion exists to convert the density distribution (6) to a phase space density distribution. We simply assume the distribution function to be

$$f = \frac{\rho(\varpi, z)}{(2\pi)^{3/2} \sigma_\varpi^2 \sigma_z} \exp\left[-\frac{1}{2}\left(\frac{\nu_\varpi^2}{\sigma_\varpi^2} + \frac{\nu_z^2}{\sigma_z^2}\right)\right]. \tag{22}$$

where $\sigma_z$ and $\sigma_\varpi$ are the velocity dispersions in the direction of the major axis and in the plane perpendicular to it, respectively. The spatial dependence of the velocity dispersions is evaluated by numerically solving the Jeans' equations in cylindrical coordinates under assumptions of axisymmetry. The required potential $\phi$ is calculated analytically in the usual way used for the ellipsoidal bodies (see e.g. BT87, pp. 49-62) (the distribution function obtained in this way is clearly not self-consistent, being correct only to the second moment, but the model results in a sufficiently stable initial equilibrium). This galaxy is let loose in the cluster to move in a circular orbit. We choose the initial conditions for the galaxy to be such that its major axis is oriented at an angle of $45°$ to the line joining the center of mass of the galaxy and the center of the cluster, and in the plane perpendicular to the velocity of the center of mass.

For all the simulations we use the Barnes & Hut (1987) code with minor modifications and the addition of a subroutine that calculates the force due to the cluster. The number of particles is fixed to $N = 32768$, and the equivalent simulation time is chosen to be shorter than the two-body relaxation time of the models. We do not use the analytical approximation to the tidal field, but the true field of the cluster. In this way we also take the nonlinear part of the external field into account. All the times reported in the figures showing the results of the numerical simulations are expressed in *simulation time units*, $(t_{\text{num}})$, with $t_{\text{num}} \simeq 0.5 t_{\text{dyn}}$ and $t_{\text{num}} \simeq 0.33 t_{\text{dyn}}$ for the spherical and the prolate models, respectively.

**Figure 8.** The isodensity contours of the projected galaxy density in the $\xi_1 - \xi_3$ plane and in the $\xi_1 - \xi_2$ plane are shown respectively in the lower and upper panels. The galaxy is in circular orbit at $\widetilde{r}_{\rm g} = 0.5$. Time is expressed in simulations units, and the space between two ticks corresponds to 4.25 kpc.

## 5 RESULTS

We have converted the end-products of the simulations into a density distribution smoothed with a Gaussian described by a smoothing length equal to about 1.5 times the force softening parameters of the code. In the simulation units the force softening is 0.2 while the boxes in Figs. (5)–(9) are 10, that corresponds to 85 kpc. The space between two ticks is 4.25 kpc. If the mass-to-light ratio is constant then the shape of the projected mass distribution will be the same as the projected light distribution. After projection we study the model surface density in the same way the observers study the surface brightness distribution, using IRAF, one of the most used image processing packages. The projected density distribution of models is fitted using 11 isophotes, and the magnitude difference between two consecutive isophotes is equal to 0.5 mag. The intensity difference from the center to the outer isophote in the figures is 5.5 mag., and using (5) and (8) this result in approximately a radial range of $3.5\,R_{\rm e}$ for the spherical model and $3.85\,l_{\rm e}$ for the prolate model, in the projection plane containing the major axis.

### 5.1 Shape modifications

We ran four simulations of an initially spherical galaxy in a cluster. The model is placed on two radial and two circular orbits. One of the radial orbits had its apocenter inside the cluster core radius ($0.75 r_{\rm c}$) while the other had it outside ($1.5 r_{\rm c}$). Similarly, one of the circular orbit lay inside the cluster core radius ($0.75 r_{\rm c}$, at which the circular orbital time in the cluster is $T_{\rm orb} \simeq 7\,10^8$ yr) while the other outside ($1.5 r_{\rm c}$, with $T_{\rm orb} \simeq 10^9$ yr).

We start by showing the case of the galaxy in the radial orbit always inside $r_{\rm c}$. In the lower panels of Fig. 5, the isodensity contours of the projected density in the $\xi_1 - \xi_3$ plane at different simulation times are shown. The galaxy orbit is along the $\xi_1$ axis, and at the beginning of the numerical simulation the galaxy was kept stationary at the distance of $0.75 r_{\rm c}$ for $20 t_{\rm num}$ as discussed in Section 4, and then released at zero velocity. In the upper panels of Fig. 5 the same galaxy at the same time is shown, but projected on the $\xi_1 - \xi_2$ plane. It is evident that the galaxy ellipticity is not affected by the CTF. The results of the simulation of a galaxy in a radial orbit with apocenter at a distance of 1.5 times the cluster core radius are essentially equally negative. This is clear from the final ellipticities shown in Table 1. Figure 4 shows that both the models in radial orbits cross a region of high radial non-adiabaticity, so we can not expect a significant effect in the shape of the galaxy. Let us then turn to the investigation of galaxies in circular orbits as possibly a constant CTF can have a stronger effect in deforming the galaxy.

Figure 6 shows the evolution of the spherical galaxy in circular orbit at $\widetilde{r}_{\rm g} = 0.75$. As in the previous cases the galaxy is initially relaxed for the same amount of time, and then released in circular orbit with the initial velocity corresponding to the position of $r_{\rm g}$. The galaxy is allowed to complete $\sim 0.8$ orbits around the cluster center. It is evident that in this case as well the deformation induced on the galaxy is negligible.

**Figure 9.** The isodensity contours of the projected galaxy density in the $\xi_1 - \xi_3$ (bottom panels) and $\xi_1 - \xi_2$ (top panels) planes are shown. The galaxy is in circular orbit at $\widetilde{r}_{\rm g} = 1.5$. Time is expressed in simulations units, and the space between two thicks corresponds to 4.25 kpc.

In Fig. 7 the projected isodensity contours of the model in circular orbit at $\widetilde{r}_{\rm g} = 1.5$ are shown. The lower four panels are the projected density distributions in the $\xi_1 - \xi_3$ plane and the upper four panels the projections in the $\xi_1 - \xi_2$ plane. The model is initially relaxed and released as in the previous simulation. This is the most favorable case for the tidal forces to produce an effect on the shape of the galaxy. So the simulation was continued for 75 $t_{\rm dyn}$, i.e. $\sim 1.7$ orbits around the cluster center. But in this case as well there is no indication of any modification of the galaxy shape.

So, even in the case of circular orbits, i.e. a situation intrinsically stationary, the effect of the CTF on the model shape is not observable. In Table 1 the maximum and minimum ellipticities, defined as usual $\epsilon = 1 - b/a$, are shown for the projected models in two different planes. The particular model is indicated by the labels that refer to the orbit (i.e. CI$_{12}$=model in circular orbit inside the cluster core radius, projected in the $\tilde{\xi}_1 - \tilde{\xi}_2$ plane). We conclude that we do not expect any systematic shape induced difference from cluster Es and field Es due to the CTF.

### 5.2 Isophotal alignment

We describe now the evolution of an initially prolate galaxy with its major axis placed in the $\xi_1 - \xi_3$ plane and inclined at $45°$ to the orbital $\xi_1 - \xi_2$ plane. The orbit of the galaxy center of mass is circular, with $\widetilde{r}_{\rm g} = 0.5$ ($T_{\rm orb} \simeq 6.3\,10^8$ yr), i.e. inside the cluster core. In Fig. 8 the galaxy's density has been projected in the $\xi_1 - \xi_3$ plane (lower panels).

The galaxy is allowed to complete 2.6 orbits around the cluster center, but alignment in the radial direction is not reached, only an increase in the ellipticity is evident. In the upper panels of Fig. 8 we plot the projected density of the same simulation in the (orbital) $\xi_1 - \xi_2$ plane. The galaxy remains rounder in this plane due to the assumed initial galaxy configuration. It is clear from this figure that while there is hardly any isophotal twisting there is a substantial increase of ellipticity (which was constant with radius at the beginning of the simulation) with time in the outer isophotes while the inner isophotes stay about the initial ellipticity. This appears to be true for density projections in and perpendicular to the plane of orbit.

We move now to the simulation of an identical galaxy with the same initial conditions, the only difference with respect to the previous one being the radius of the circular orbit, now $\widetilde{r}_{\rm g} = 1.5$. Figure 9 shows the projections of the galaxy density in the $\xi_1 - \xi_3$ and $\xi_1 - \xi_2$ planes. We do observe isophotal alignment in this case, more than in the model in the inner orbit. But not very much isophotal twisting is seen. In this case as well there is evidence for the increase of ellipticity in the outer isophotes although not quite as dramatic as the case when the galaxy is in circular orbit inside the cluster core radius. Obviously the first conclusion we can draw from this is that there will be a segregation in the alignment property of elliptical galaxies in clusters. Beyond this however is the observation that there is a marked increase of ellipticity with time and this increase is more in the outer isophotes than in the inner isophotes. This is in contrast to the situation for galaxies that were initially spherical where the CTF appeared to have no effect.

**Table 1.** Maximum and minimum isophotal ellipticities of spherical model between 0.5 and $3.5R_e$. The projection plane is indicate by the subscripts and all values are reported for the last simulation time reported in the figures.

| Ellipticity | $RI_{12}$ | $RI_{13}$ | $RO_{12}$ | $RO_{13}$ | $CI_{12}$ | $CI_{13}$ | $CO_{12}$ | $CO_{13}$ |
|---|---|---|---|---|---|---|---|---|
| $\epsilon_m$ | 0.00 | 0.00 | 0.02 | 0.00 | 0.00 | 0.00 | 0.01 | 0.00 |
| $\epsilon_M$ | 0.09 | 0.09 | 0.09 | 0.07 | 0.04 | 0.07 | 0.07 | 0.08 |

## 6 CONCLUSIONS

In this paper we report the results of N-body simulations of spherical and prolate Es in a spherical rich cluster. We studied the problem of how the mean tidal field of the cluster affect the Es inside it: we search for possible galaxy shape deformations and global alignment of prolate galaxies with the cluster radial direction. A preliminary analytical study show that (i) the topological change in the CTF near $r_c$ causes galaxies on radial obits through this region to see non-adiabatic potential changes and (ii) the alignment times for prolate galaxies are much longer than their dynamical time scales. The galaxy models are placed in circular and radial orbits. The main results are (confirming the analysis in Section 3) that the shape modifications produced by the CTF on the spherical models are not detectable up to many effective radii, but a substantial major axis alignment with the cluster radial direction is reached in a few galactic dynamical times for the prolate galaxies in circular orbit outside the cluster core radius. This alignment is due to the torque induced by the tidal field on the galaxy inertia tensor and to the conversion into disordered stellar motion of the ordered rotational energy acquired by the galaxy. We suggest a possible population segregation as a function of distance of the galaxy from the cluster center. The galaxies that are outside the cluster core radius will have a greater tendency to have their major axes aligned in few internal dynamical times in the direction to the cluster center than the galaxies inside the cluster core radius. This effect allows us to reconcile a global galaxy alignment with a bottom-up galaxy formation scenario. We observe in our simulations an increase of the ellipticity of the external isophotes. This is due to the fact that the tidal force increases linearly with the distance from the center of galaxy, while the ratio between the self-gravity of the galaxy with respect to the CTF decreases in the outer galaxy regions as the third power of the distance from the galactic center. We also found no systematic isophotal twisting of the galaxies, and this is consistent with the observational results of Fasano & Bonoli (1989), that exclude the possibility of CTF being major contributor to the isophotal twisting in Es.

## ACKNOWLEDGMENTS

We are grateful to Jerry Ostriker and David Spergel for useful discussions, to Ralf Bender, George Djorgovsky and Massimo Stiavelli for comments on the observational section, and to Joshua Barnes for kindly providing the code used for the numerical simulations. One of us (L.C.) acknowledge the warm hospitality of the Princeton University Observatory, where he was supported by the NSF Grant AST 9108103. S.N.D. acknowledges the support of the NSF Grants AST88-58145 and AST91-17388. This work was partially supported by the EEC contract No. CHRX-CT92-0033 and by the Italian Ministry of Research (MURST).

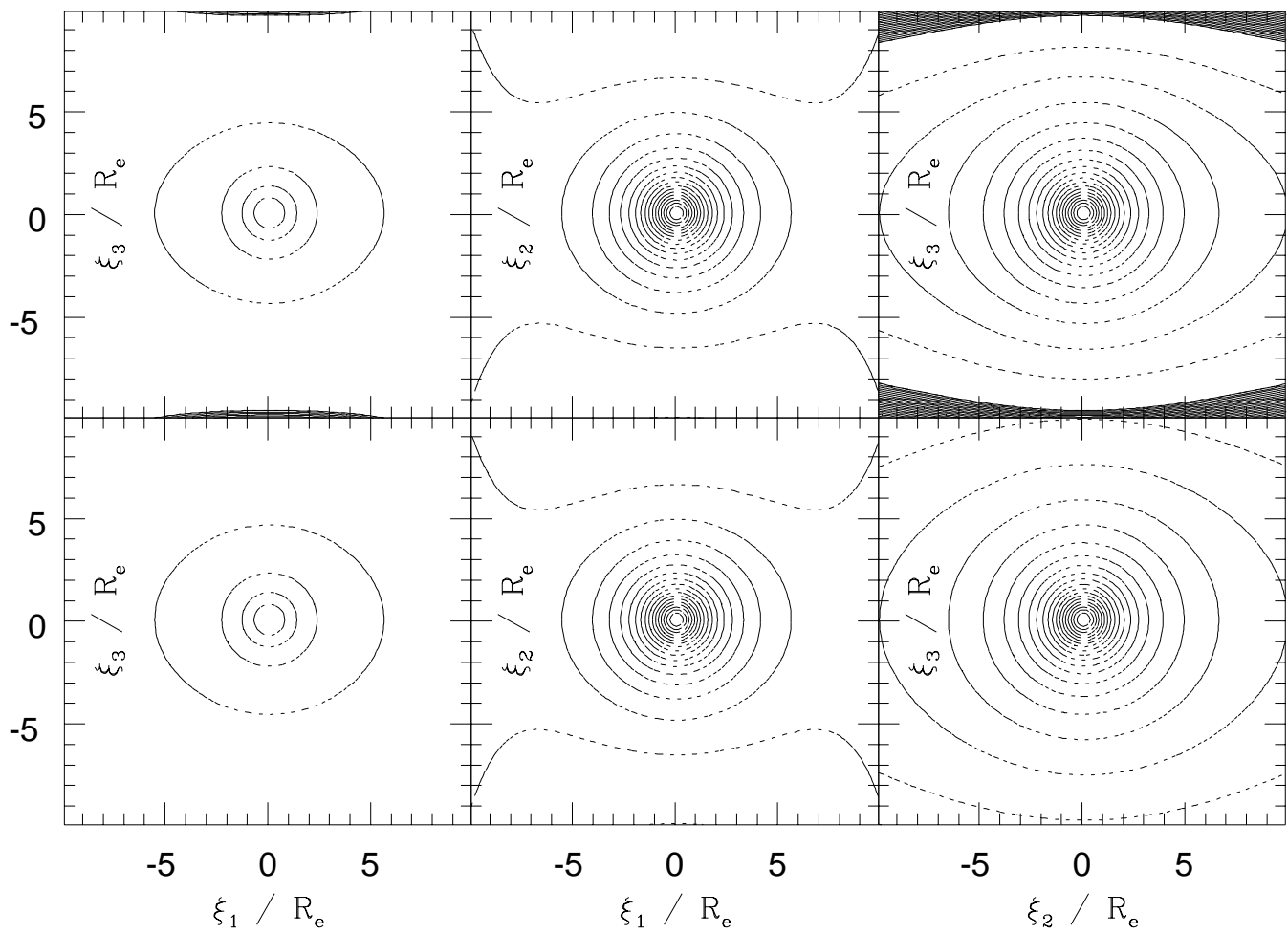

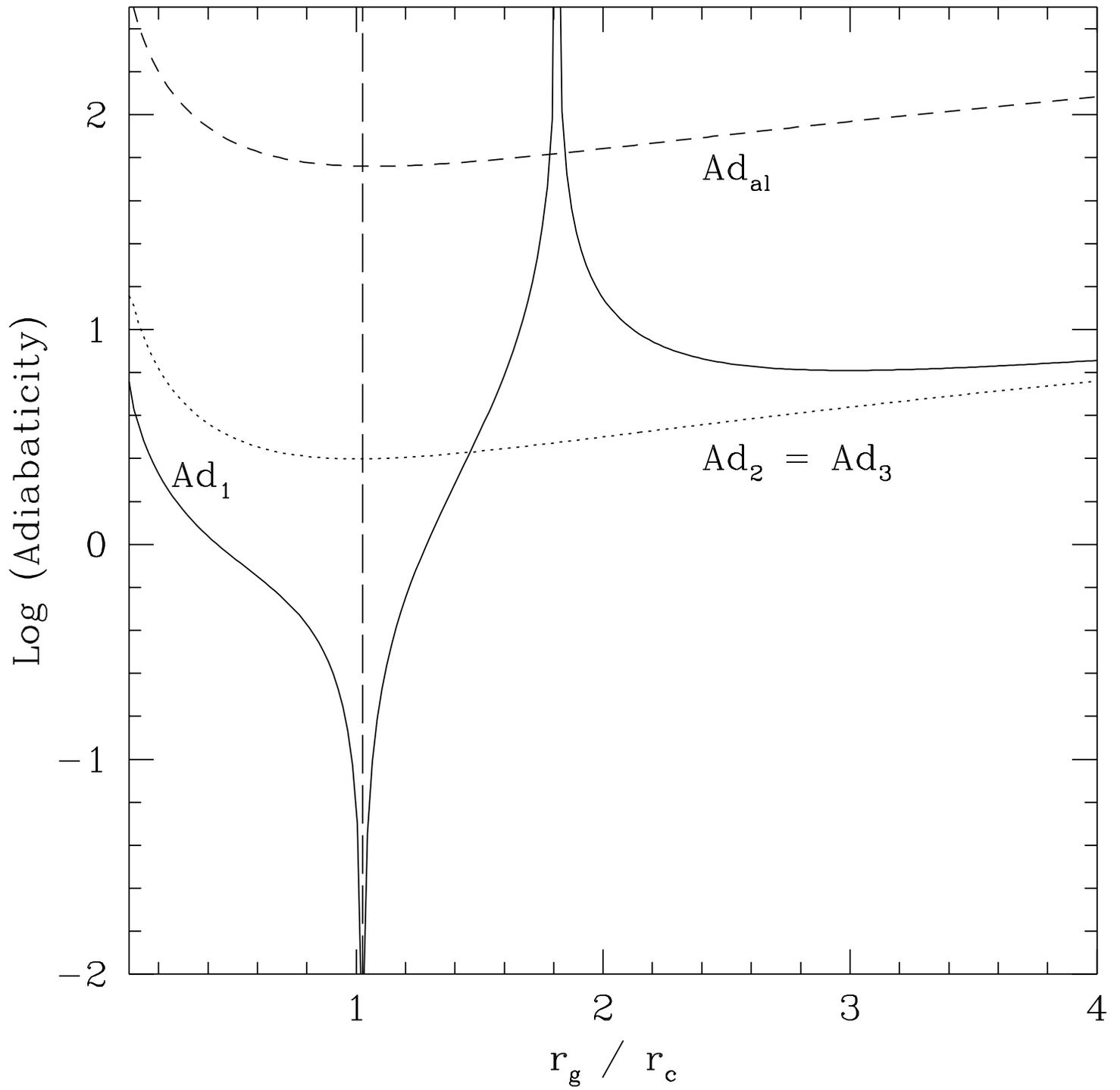

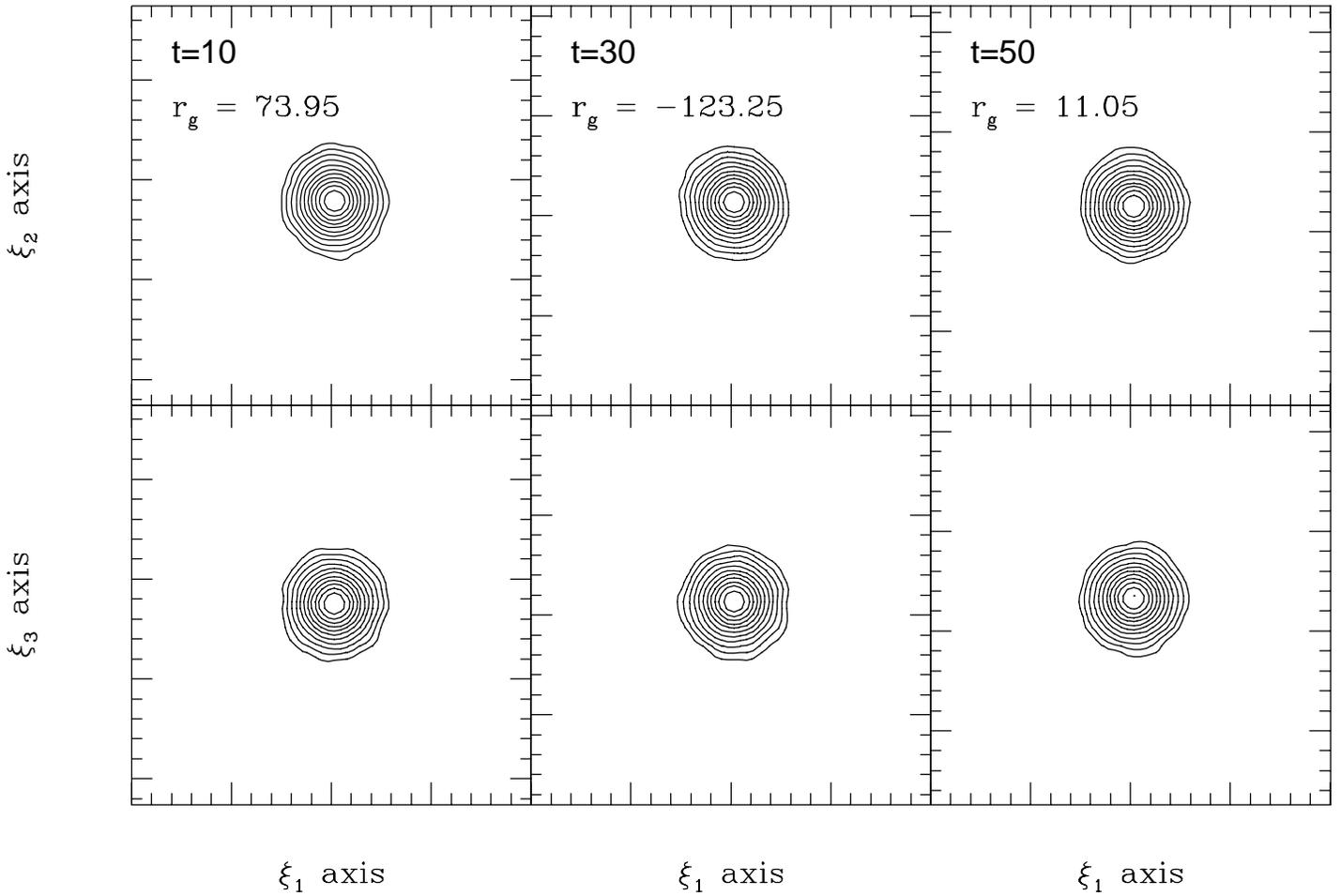

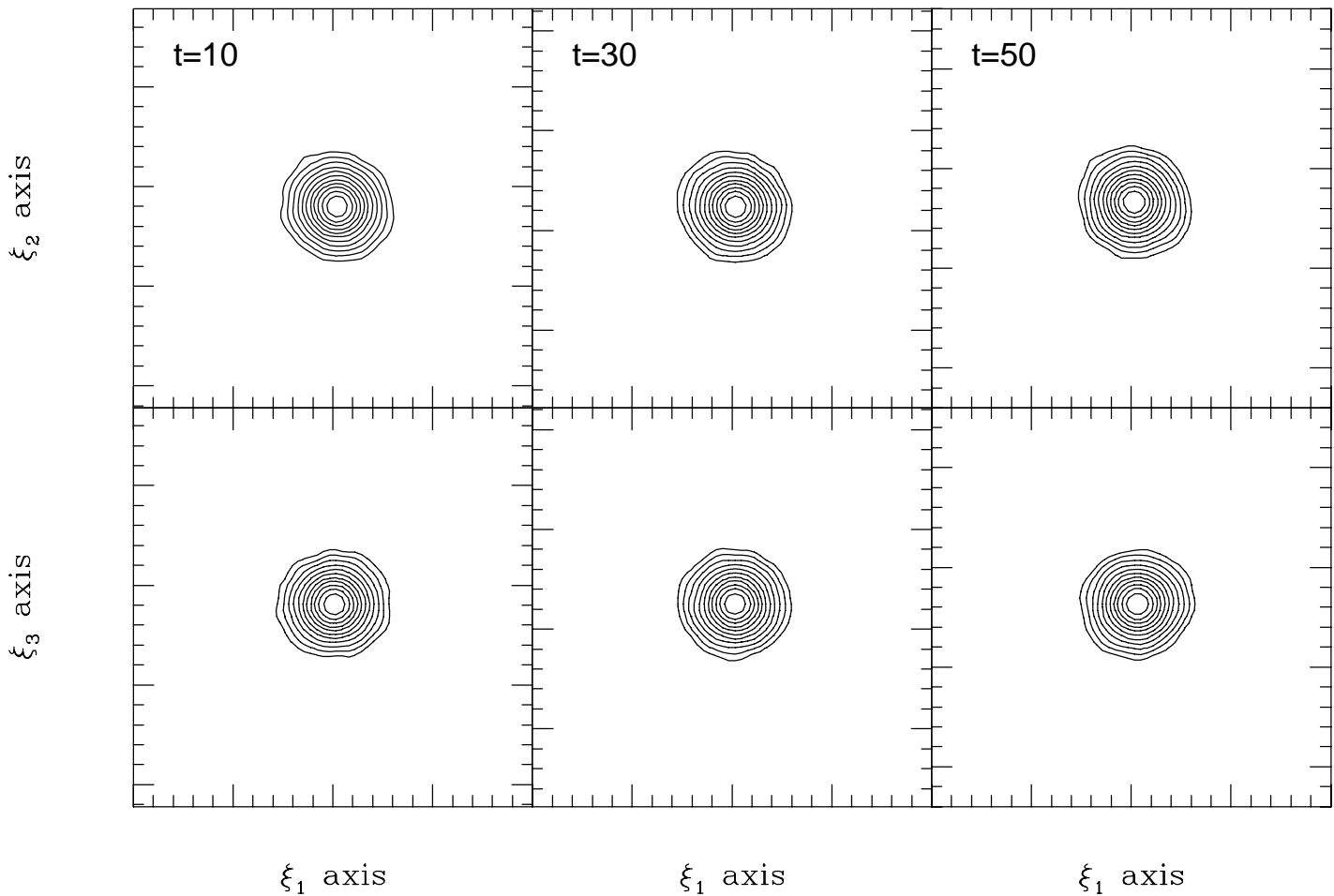

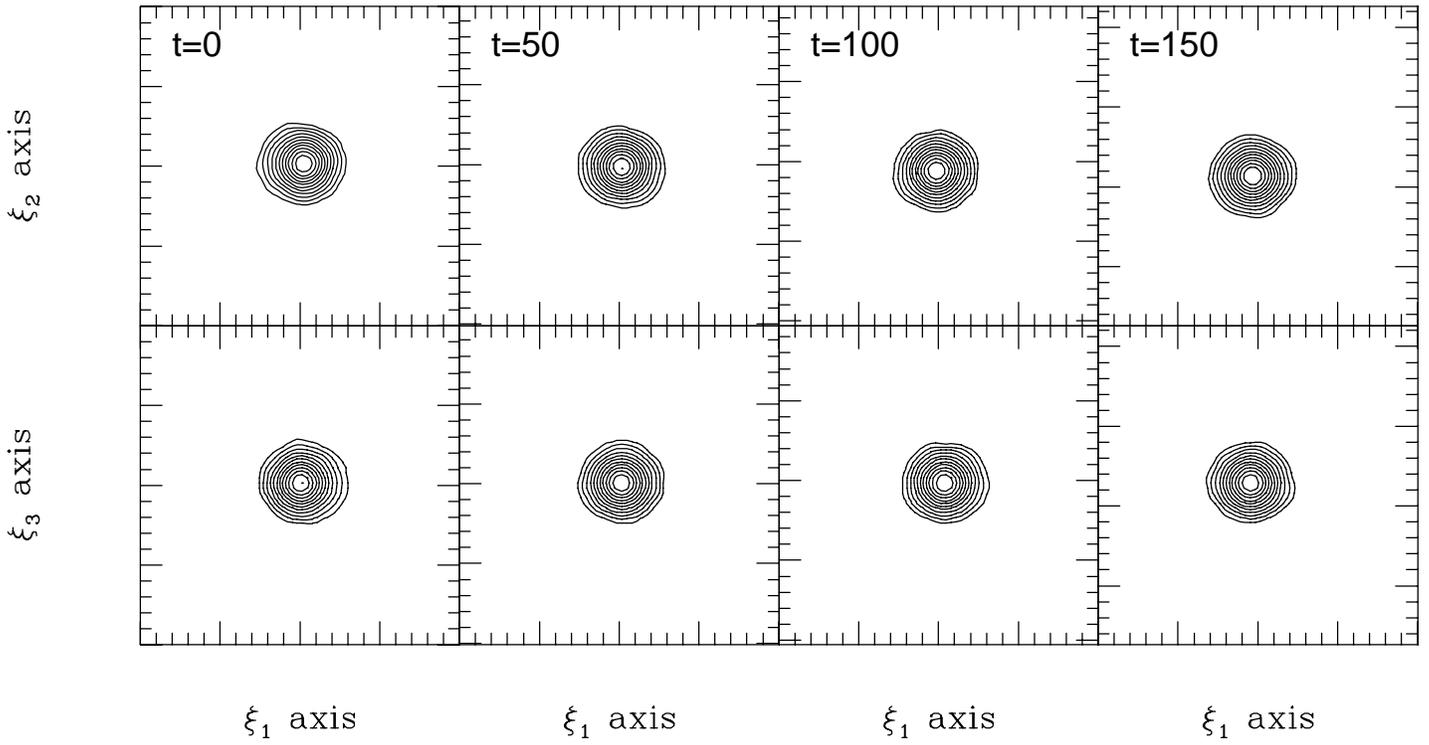

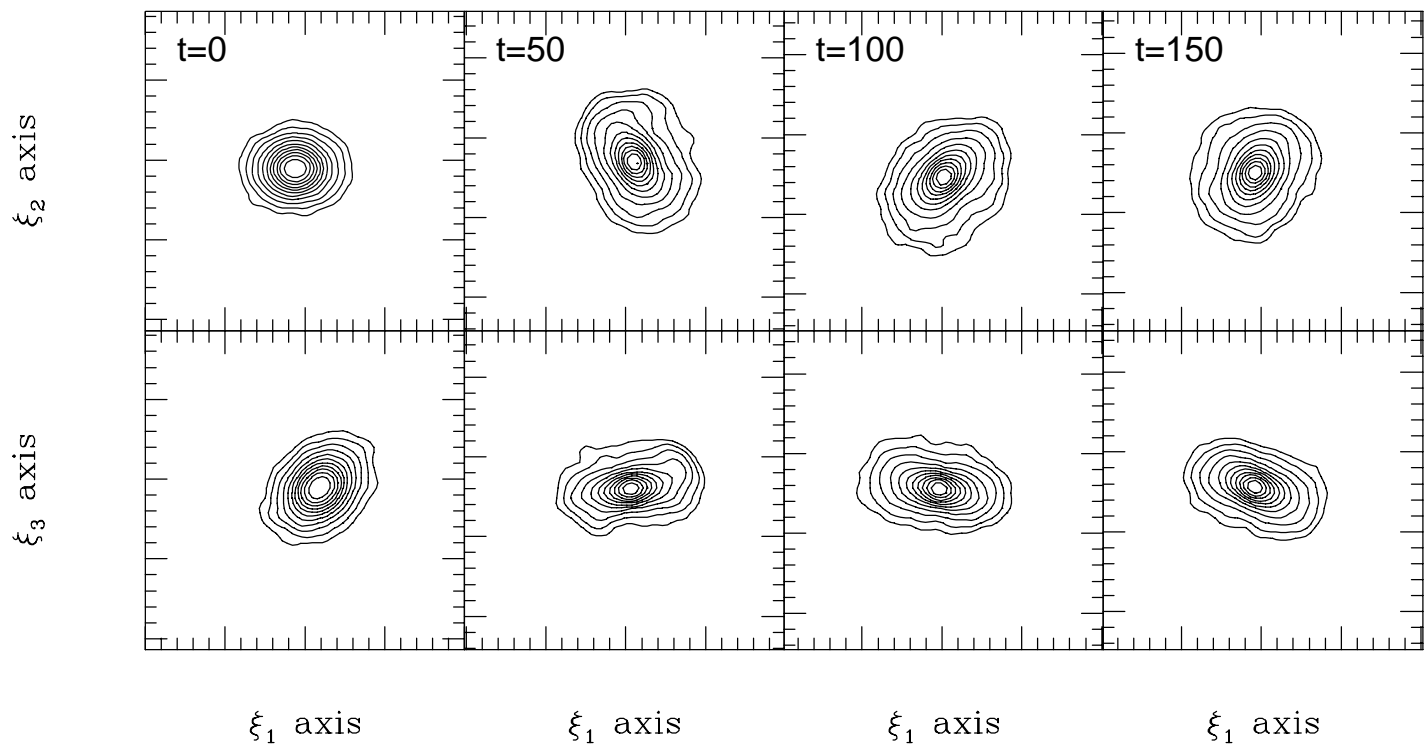

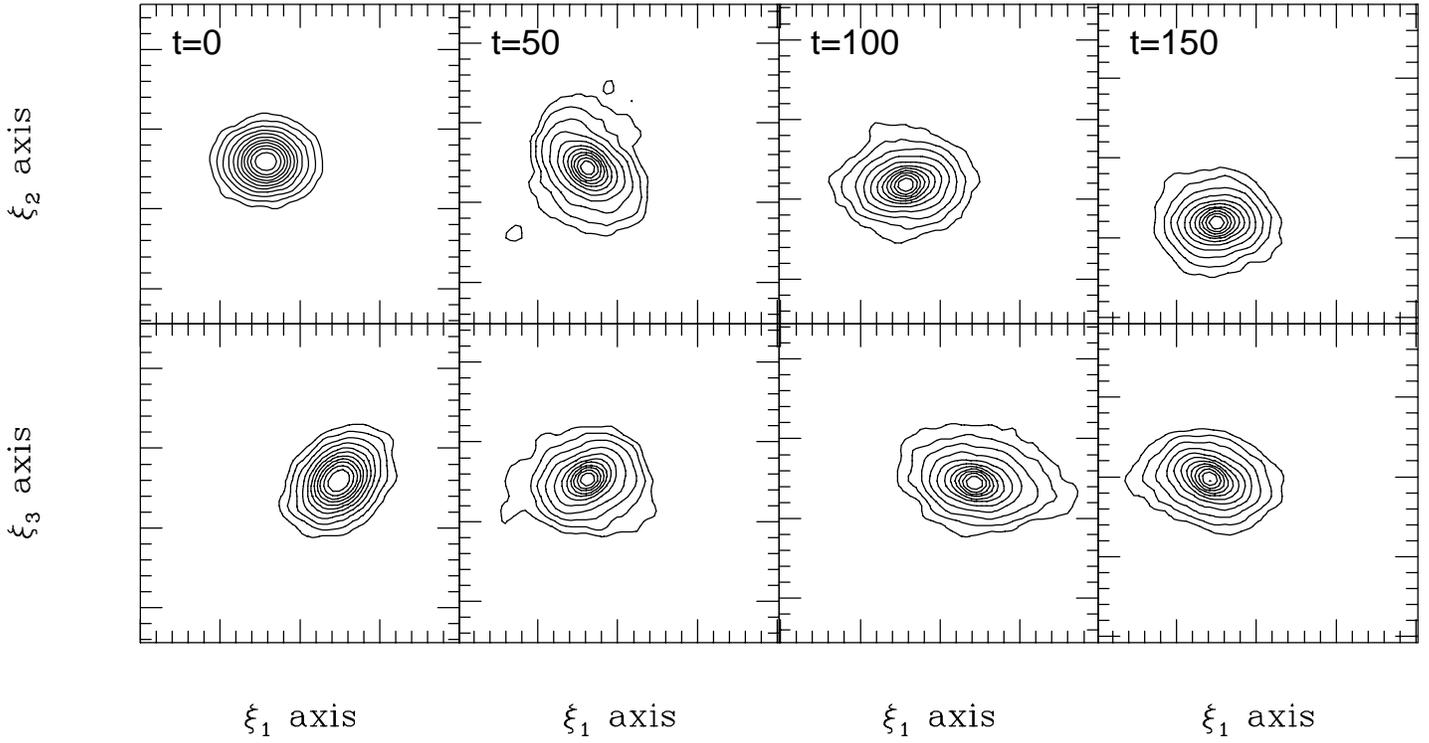